\documentclass[12pt]{article}
\usepackage[a4paper, total={7in, 10in}]{geometry}
\usepackage[parfill]{parskip}
\usepackage{physics, tensor, float, subcaption}
\usepackage{graphicx}
\graphicspath{ {Plots/} }
\usepackage{jhep-mod}
\usepackage{bm}
\usepackage{soul}
\usepackage{amssymb,amsmath,amsthm}
\usepackage{mathrsfs}
\usepackage[utf8]{inputenc}
\usepackage{enumerate}
\usepackage{bigints}
\usepackage{xcolor}
\usepackage{appendix}
\usepackage{graphicx}
\usepackage{float}
\usepackage{tikz}
\usepackage{setspace}
\usepackage{cancel}
\usepackage{array}
\usepackage{tabulary}
\usepackage{doi}
\definecolor{purple}{rgb}{1,0,1}
\definecolor{lime}{HTML}{A6CE39} 

\newcommand{\blue}[1]{{\color{blue} #1}}

\definecolor{lime}{HTML}{A6CE39}
\newcommand{\orcidicon}{%
	\begin{tikzpicture}
	\draw[lime, fill=lime] (0,0) 
		circle [radius=0.16] 
		node[white] {{\fontfamily{qag}\selectfont \tiny ID}};
	\draw[white, fill=white] (-0.0625,0.095) 
		circle [radius=0.007];
	\end{tikzpicture}
	\hspace{-5mm}
}
\newcommand\orcidChris{{\href{https://orcid.org/0009-0000-3953-0461}{\orcidicon}}}
\newcommand\orcidMatt{{\href{https://orcid.org/0000-0003-1088-6485}{\orcidicon}}}

\renewcommand{\O}{\mathcal{O}}

\newcommand{\be}{\begin{equation}}
\newcommand{\ee}{\end{equation}}

\def\O{{\mathcal{O}}}
\def\sign{\mathrm{{sign}}}
\begin{document}
\newcommand{\arXiv}[1]{arXiv:\href{https://arxiv.org/abs/#1}{\color{blue}#1}}

\title{\vspace{-25pt}\huge{
\blue{Pressure profile bounds from relaxing the TOV equation}
}}


\author{\Large Christopher Simmonds\!\orcidChris\;}
\emailAdd{chris0simmonds@gmail.com}
\author{\!\!and \Large Matt Visser\!\orcidMatt$^{\dagger}$}
\emailAdd{matt.visser@sms.vuw.ac.nz}
\affiliation{School of Mathematics and Statistics, Victoria University of Wellington, \\
\null\qquad PO Box 600, Wellington 6140, New Zealand.}
\affiliation{$^\dagger$ Corresponding author.}
\renewcommand{\arXiv}[1]{arXiv:\href{https://arxiv.org/abs/#1}{\color{blue}#1}}
\def\L{{\mathcal{L}}}
\def\theta{\vartheta}
\def\phi{\varphi}

\abstract{ \\
We develop several new and quite general bounds on the internal pressure profiles of general relativistic perfect fluid spheres --- based on various ways of relaxing the TOV system of ODEs to obtain several distinct differential inequalities.
There is, as usual,  a trade-off between strength of the bound, weakness of the input assumptions, 
and tractability of the analysis.
Specifically we shall develop several straightforward but nontrivial bounds that variously depend only on 
the positivity of density, the boundedness of density, the monotonicity of density, or the monotonicity of the average density.  
We shall carefully place these new bounds within the historical framework of previous efforts in this regard, and explore the ways in which they are inter-related. 


 
\bigskip
\noindent
{\sc Date:}  Wednesday 12 August 2026; \LaTeX-ed \today.

 
\bigskip 
\bigskip
\hrule\hrule\hrule

}

\maketitle
\def\tr{{\mathrm{tr}}}
\def\diag{{\mathrm{diag}}}
\def\cof{{\mathrm{cof}}}
\def\pdet{{\mathrm{pdet}}}
\def\QED{ {\hfill$\Box$\hspace{-25pt}  }}
\def\d{{\mathrm{d}}}
\def\sign{\hbox{sign}}
\def\theta{\vartheta}
\def\phi{\varphi}

\parindent0pt
\parskip7pt

\clearpage
\null
\vspace{-75pt}
\section{Introduction}

There is a long-standing tradition of using both Newtonian gravity and Einstein gravity to place 
bounds on the internal pressure profile of perfect fluid spheres~\cite{Chandrasekhar:1942, Tolman:1934, Tolman:1939, Oppenheimer:1939, Buchdahl:1959, Bondi:1964, Buchdahl:1966, Kovetz:1968, Kovetz:1969, Islam:1969, Islam:1970, Forrester:1971, Guven:1994, Guven:1997, Guven:1999, Delgaty:1998, Martin:2003, Boonserm:2005a, Boonserm:2005b, Boonserm:2006a, Boonserm:2006b, Martin:2006, Faber:2006, Boonserm:2007a, Boonserm:2007b, Mak:2013, Breu:2016, Mottola:2023, Simmonds:2025, Reintjes:2025, Maier:2026}\\ --- as a zeroth order approximation to understanding idealized stellar structure. Many of these bounds depend on  relatively strong monotonicity assumptions on the volume averaged density
\begin{equation}
\bar\rho(r) = {m(r)\over{4\pi\over 3} r^3} 
= {\int_0^r 4\pi \rho(\bar r) \bar r^2 d\bar r \over{4\pi\over 3} r^3}
= {\int_0^r 4\pi \rho(\bar r) \bar r^2 d\bar r \over \int_0^r 4\pi  \bar r^2 d\bar r},
\end{equation}
such as assuming $d\bar\rho(r)/dr \leq 0$, or even the slightly stronger $d\bar\rho(r)/dr < 0$. 

While in some specific situations such a condition on the volume averaged density might be physically plausible, its validity is by no means universally justifiable. In view of this, we shall also consider several weaker conditions on the density, and derive a number of simple bounds on the pressure profile under these weakened conditions. 
We also present an improved bound based on monotonicity of average density.

Specifically, we shall consider the following four conditions on the density:
\begin{itemize}
\item Positivity of density, without any monotonicity condition.
\item Boundedness of density, without any monotonicity condition.
\item Monotonicity of density.
\item Monotonicity of average density.
\end{itemize}
We shall soon see that even the weakest of these four conditions will still tell us something interesting.
We shall also carefully explore the manner in which these input conditions are inter-related, and the manner in which the resulting derived bounds are inter-related.

\bigskip

\clearpage
\section{Fundamentals: The TOV system of ODEs}
 
We shall work within the framework of  the well-known TOV system (Tolman--Oppenheimer--Volkoff system) of ordinary differential equations (ODEs). Namely:
\begin{equation}
{d p(r) \over dr} = -{[\rho(r)+p(r)]\; [m(r)+4\pi p(r) r^3]\over r^2(1-2m(r)/r)};
\end{equation}
\begin{equation}
{d m(r) \over dr} = 4\pi \rho(r) r^2;
\end{equation}
 \begin{equation}
\rho(r) = f(p(r)).
\end{equation}
See for instance references~\cite{Tolman:1934, Tolman:1939, Oppenheimer:1939, Buchdahl:1959, Bondi:1964, Buchdahl:1966, Kovetz:1968, Kovetz:1969, Islam:1969, Islam:1970, Forrester:1971, Guven:1994, Guven:1997, Guven:1999, Delgaty:1998, Martin:2003, Boonserm:2005a, Boonserm:2005b, Boonserm:2006a, Boonserm:2006b, Martin:2006, Faber:2006, Boonserm:2007a, Boonserm:2007b, Mak:2013, Breu:2016, Mottola:2023, Simmonds:2025, Reintjes:2025, Maier:2026}, and also the book-level discussions in~\cite{Harrison:1965,  Weinberg:1972, MTW, Wald:1984, Stephani:2003, Griffiths:2009}.

Traditionally one most typically takes one of these two routes:
\begin{itemize}
\item 
Start at the origin, $r=0$.\\
Pick some central pressure $p(0)=p_c$, and set $m(0)=0$. \\
Then integrate outwards from $r=0$ until the pressure $p(r)$ drops to zero.\\
This defines the surface of the star: $p(r_s)=0$.
\item
Start at the surface of the star, pick some convenient $r_s$.\\
Set $p(r_s)=0$, and pick some mass $m(r_s)=m_s$, while satisfying $2m_s < r_s$. \\
Then integrate  inwards from $r=r_s$ until one reaches $r=0$.
\end{itemize}
A less common option is to take a third ``mixed'' route:
\begin{itemize}
\item Pick some convenient $r_*$.\\ 
Set $p(r_*)=p_*$ and $m(r_*)=m_*$, while  satisfying $2m_* < r_*$. \\
Then integrate outwards from $r=r_*$ until the pressure $p(r)$ drops to zero,\\
and also integrate  inwards from $r=r_*$ until one reaches $r=0$.
\end{itemize}
We shall adapt and modify all three of these routes in implementing the discussion, 
and proving the various theorems and corollaries, to be developed below.

\enlargethispage{20pt}
Even though the TOV system is first-order, implying that the only mathematically important boundary conditions are the values of the various fields at the starting point of the integration, for comparison purposes it is sometimes useful to have on hand some extra information about higher derivatives. 

\begin{itemize}
\item
At this stage we shall only demand finiteness and differentiability of pressure $p(r)$ and density $\rho(r)$. At the origin $r=0$, the centre of the star,  this requires
\begin{equation}
p(r) = p_c + \O(r^2); \qquad \rho(r)=\rho_c + \O(r^2); \qquad
m(r) = {4\pi\over 3} \rho_c r^3 + \O(r^5). 
\end{equation}
See for instance references~\cite{Visser-Yunes} and~\cite{Arrechea:2026}.
Note that at this stage we are not (yet) demanding positivity of density or pressure.

\item 
Inserting these near-centre expressions into the TOV leads to 
\begin{equation}
p(r) = p_c - {2\pi\over 3}[\rho_c+p_c][\rho_c+3p_c] r^2 + \O(r^4).
\end{equation}

Thence, in addition to the boundary condition $\left.{p(r)}\right|_{r=0} = p_c$, we obtain the following two constraints
\begin{equation}
\label{E:d^2p/dr^2-at-centre}
\left.{dp(r)\over dr}\right|_{r=0}= 0;
\qquad\qquad
\left.{d^2p(r)\over dr^2}\right|_{r=0} =  -{4\pi\over 3}[\rho_c+p_c][\rho_c+3p_c].
\end{equation}
(With a bit more effort, the central value of the  fourth derivative $\left.{d^4p(r)/ dr^4}\right|_{r=0}$ can also be extracted. This is discussed in the Appendix.)

\item
Similarly, at the surface of the star, in addition to the boundary condition  $p(r_s)=0$ one also knows that 
\begin{equation}
\left.{dp(r)\over dr}\right|_{r_s} = -{\rho_s m_s\over r_s^2(1-2m_s/r_s)}.
\end{equation}
\item
Finally, if one starts the integrations at some $r_*\in(0,r_s)$, then in addition to the boundary conditions $p(r_*)=p_*$ and $m(r_*)= m_*$, one also knows that
\begin{equation}
\label{E:slope-at-star}
\left.{dp(r)\over dr}\right|_{r_*} = -{[\rho_*+p_*][ m_*+4\pi p_* r_*^3] \over r_*^2(1-2m_*/r_*)}.
\end{equation}
\end{itemize}
These extra pieces of information are valid under extremely general conditions and will prove useful when comparing various models to be developed below.

In a totally different direction, note that the corresponding Newtonian system of ODEs is simply
\begin{equation}
{d p(r) \over dr} = -{\rho(r)\; m(r) \over r^2};
\qquad
{d m(r) \over dr} = 4\pi \rho(r) r^2;
\qquad
\rho(r) = f(p(r)).
\end{equation}
This Newtonian system is considerably easier to analyze~\cite{Chandrasekhar:1942}. 

 \bigskip

\clearpage
\section{Theorem 1 (positive density) }

We shall start with the extremely weak constraint of only assuming positive density: $\rho(r)>0$ for $r \geq 0$.
Then, since this guarantees $m(r)>0$ for $r>0$, we see
\begin{equation}
{[\rho(r)+p(r)]\; [m(r)+4\pi p(r) r^3]\over r^2(1-2m(r)/r)} > {p(r)\; [4\pi p(r) r^3]\over r^2}= 4\pi p(r)^2 r.
\end{equation}
Thus the entire TOV system of ODEs collapses to a single differential inequality
\begin{equation}
\label{E:dineq1}
{d p(r) \over dr} < -4\pi p(r)^2 r; \qquad (r>0).
\end{equation}
(This differential inequality, while extremely simple,  is still sufficient to capture the notion of ``regeneration of pressure'' that is one of the hallmarks of Einstein gravity.) Rearranging, we see
\begin{equation}
\label{E:dineq1b}
{d\over dr}\left( {1\over p(r)} - 2\pi r^2 \right) > 0; \qquad (r>0).
\end{equation}
From this we rapidly deduce the following.

{\bf Theorem 1:}\\
Suppose we know the pressure $p_*=p(r_*)$ at some point $r_*$.\\
Using only the fact that density is positive inside the star, $\rho(r)>0$, we have
\begin{equation}
\label{E:T1a}
p(r) < {p_*\over1+2\pi p_* [r^2-r_*^2]}; \qquad (r>r_*).
\end{equation}
\begin{equation}
\label{E:T1b}
p(r) > {p_*\over1+2\pi p_* [r^2-r_*^2]}; \qquad (r<r_*).
\end{equation}

\enlargethispage{25pt}
{\bf Proof:} \\
First, suppose $r>r_*$, then integrating the differential inequality (\ref{E:dineq1b}) upwards, from $r_*$ up to some $r>r_*$,  we see
\begin{equation}
\left( {1\over p(r)} - 2\pi r^2 \right) > \left( {1\over p_*} - 2\pi r_*^2 \right) ; \qquad (r>r_*).
\end{equation}
Rearrange to obtain (\ref{E:T1a}).

Second, suppose $r<r_*$, then integrating the differential inequality (\ref{E:dineq1b})  upwards, but now from from $r$ up to $r_*$, we see
\begin{equation}
\left( {1\over p_*} - 2\pi r_*^2 \right) > \left( {1\over p(r)} - 2\pi r^2 \right) ; \qquad (r<r_*).
\end{equation}
Rearrange this  to obtain (\ref{E:T1b}). \\
(Formally speaking, only the direction of the inequality has changed.)
\hfill{$\Box$}

\clearpage
{\bf Comment:} \\
This is certainly not the most stringent of bounds, 
but  it is extremely simple to state and to prove.
Furthermore, under the stated hypothesis the bound is in fact optimal --- for instance it is easy to check that Schwarzschild's zero density star saturates these bounds, converting them into an equality~\cite{revisiting}. 

{\bf Corollary 1a:}\\
\leftline{Under the conditions of Theorem 1, the central pressure $p(0)=p_c$ is bounded by}
\begin{equation}
p_c> {p_*\over1-2\pi p_* r_*^2}.
\end{equation}

{\bf Proof:} Obvious. \hfill{$\Box$}

{\bf Corollary 1b:}
Suppose the central pressure, $p(0)=p_c$, is finite and positive.\\
Using only the fact that density is positive inside the star, $\rho(r)>0$, we have
\[
p(r) < {p_c\over1+2\pi p_c r^2}; \qquad (r>0).
\]
It is easy to check that Schwarzschild's zero density star saturates this bound, \\ converting it into an equality~\cite{revisiting}. 

In particular, independent of the specific value of $p_c$, we have
\[
p(r) < {1\over2\pi r^2}; \qquad (r>0).
\]

{\bf Proof:} Obvious. \hfill{$\Box$}

Now let us see what we can do with stronger input assumptions.

\bigskip
 
\clearpage
\section{Theorem 2 (bounded positive density)}

Let us now see what we can do by merely assuming the density is aways bounded between some positive surface density $\rho_s = \rho(r_s)$ and the larger central density $\rho_c=\rho(0)$, \emph{but without making any monotonicity assumptions}. 
That is
\begin{equation}
\rho_c > \rho(r) >  \rho_s >0 \qquad \hbox{for} \qquad r\in(0,r_s).
\end{equation}
First recall that $m(r) = \int_0^r 4\pi \rho(\bar r) \bar r^2 d \bar r$. Now define $\hat m(\rho_0,r) = {4\pi\over3} \rho_0 r^3$. \\
Then under the stated conditions we have
\begin{equation}
\hat m(\rho_c, r) > m(r) > \hat m(\rho_s, r) >0 \qquad \hbox{for} \qquad r\in(0,r_s).
\end{equation}
Thence
\begin{equation}
{[\rho(r)+p(r)]\; [m(r)+4\pi p(r) r^3]\over r^2(1-2m(r)/r)} >
{[\rho_s +p(r)]\; [\hat m(\rho_s, r) +4\pi p(r) r^3]\over 
r^2(1-2\hat m(\rho_s, r) /r)};
\end{equation}
and
\begin{equation}
{[\rho(r)+p(r)]\; [m(r)+4\pi p(r) r^3]\over r^2(1-2m(r)/r)} < 
{[\rho_c +p(r)]\; [\hat m(\rho_c, r) +4\pi p(r) r^3]\over 
r^2(1-2\hat m(\rho_c, r) /r)},
\end{equation}
implying the two differential inequalities 
\begin{equation}
\label{E:diff-ineq1}
{dp(r)\over dr} <
- {[\rho_s +p(r)]\; [\hat m(\rho_s, r) +4\pi p(r) r^3]\over 
r^2(1-2\hat m(\rho_s, r) /r)};
\end{equation}
and 
\begin{equation}
\label{E:diff-ineq2}
{dp(r)\over dr} > 
- {[\rho_c +p(r)]\; [\hat m(\rho_c, r) +4\pi p(r) r^3]\over 
r^2(1-2\hat m(\rho_c, r) /r)}.
\end{equation}
Now the right hand side (RHS) of these two differential inequalities is exactly the quantity one would expect to see when analyzing Schwarzschild's constant density star~\cite{Wald:1984, Schwarzschild-star, Schwarzschild-star-translation}. (For either $\rho_0\to \rho_s$ or $\rho_0 \to \rho_c$ respectively.) Based on this observation, with a little experimentation one finds that for \emph{any} constant density $\rho_0$, and \emph{any} arbitrary dimensionless constant $K_0$, the quantity
\begin{equation}
\hat p(\rho_0, K_0, r) = \rho_0 \left( \sqrt{1-2\hat m(\rho_0, r) /r} - K_0 \over 3 K_0 -  \sqrt{1-2\hat m(\rho_0, r) /r}\right),
\end{equation}
satisfies the differential \emph{equality}
\begin{equation}
{d\hat p(\rho_0,K_0, r)\over dr} =
- {[\rho_0 +\hat p(r)]\; [\hat m(\rho_0, r) +4\pi \hat p(r) r^3]\over 
r^2(1-2\hat m(\rho_0, r) /r)}.
\end{equation}
Explicitly we have
\begin{equation}
\hat p(\rho_0, K_0, r) = \rho_0 \left( \sqrt{1-{8\pi\over3} \rho_0 r^2} - K_0 \over 3 K_0 -  \sqrt{1-{8\pi\over3} \rho_0 r^2}\right).
\end{equation}

To keep $\hat p(\rho_0,K_0, r)$ both finite and positive throughout the star one should restrict attention to the parameter region 
\begin{equation}
K_0 \in\left({1\over3}, \sqrt{1-{8\pi\over3} \rho_0 r_s^2}\right).
\end{equation}
It is for some purposes sufficient to merely keep  $\hat p(\rho_0,K_0, 0)$ finite and positive at the centre of the star, merely
restricting attention to the somewhat wider parameter region $K_0 \in\left({1\over3}, 1\right)$.
The otherwise arbitrary dimensionless constant $K_0$ will soon be used to fit various appropriate boundary/\-initial/\-final conditions.
By now considering the quantities $p(r)-\hat p(\rho_0,K_0,r)$ we can formulate the following argument.

Consider  for instance the upper bound based on the surface density $\rho_s$:
\begin{equation}
{d\over dr} p(r) < 
- {[\rho_s +p(r)]\; [\hat m(\rho_s, r) +4\pi p(r) r^3]\over  r^2(1-2\hat m(\rho_s, r) /r)} 
\end{equation}
whence
\begin{eqnarray}
{d\over dr} [p(r)-\hat p(\rho_s,K_0,r)] < 
&-& {[\rho_s +p(r)]\; [\hat m(\rho_s, r) +4\pi p(r) r^3]\over  r^2(1-2\hat m(\rho_s, r) /r)} \nonumber\\
&+& {[\rho_s +\hat p(r)]\; [\hat m(\rho_s, r) +4\pi \hat p(r) r^3]\over  r^2(1-2\hat m(\rho_s, r) /r)} \nonumber\\
&=& -{[p(r)-\hat p(r)][\hat m(\rho_s,r)+4\pi\rho_s r^3] + 4\pi [p(r)^2 - \hat p(r)^2] r^3\over  r^2(1-2\hat m(\rho_s, r) /r)} \nonumber\\
&=& -{[p(r)-\hat p(r)][4 \hat m(\rho_s,r)+ 4\pi [p(r) + \hat p(r)] r^3]\over  r^2(1-2\hat m(\rho_s, r) /r)} \nonumber\\
&=& - [p(r)-\hat p(r)] \times \{\hbox{something positive}\}.
\end{eqnarray}
That is, being a little more explicit about the parameters involved, we have
\begin{equation}
{d\over dr} [p(r)-\hat p(\rho_s,K_0,r)] <   - [p(r)-\hat p(\rho_s,K_0,r)] \times \{\hbox{something positive}\}.
\end{equation}
Thence, if $p(r)>\hat p(r)$ we see
\begin{equation}
{d\over dr} [p(r)-\hat p(\rho_s,K_0,r)] <   0.
\end{equation}
Similarly  if $p(r)<\hat p(r)$ we see
\begin{equation}
{d\over dr} [p(r)-\hat p(\rho_s,K_0,r)] >   0.
\end{equation}
More prosaically we can summarize these two inequalities as
\begin{equation}
{d\over dr} \left| p(r)-\hat p(\rho_s,K_0,r)\right| \leq   0,
\end{equation}
with equality (of the derivatives) only when $p(r)=\hat p(r)$.

Similarly for the lower bound based on the central density $\rho_c$, in an entirely analogous manner, one sees that:
\begin{equation}
{d\over dr} \left| p(r)-\hat p(\rho_c,K_0,r)\right| \geq  0,
\end{equation}
with equality (of the derivatives) only when $p(r)=\hat p(r)$.

We now summarize this discussion by formulating a theorem. 

{\bf Theorem 2:}\\
Suppose merely that the density is bounded $\rho_c > \rho(r) >  \rho_s$ for $r\in(0,r_s)$.\\
Let $K_c$ and $K_s$ be arbitrary constants. Then for $r\in(0,r_s)$ we have
\begin{equation}
{d\over dr} \left| p(r) 
-  \rho_s \left( \sqrt{1-2\hat m(\rho_s, r) /r} - K_s \over 3 K_s -  \sqrt{1-2\hat m(\rho_s, r) /r}\right)\right|
 \leq 0;
\end{equation}
and
\begin{equation}
{d\over dr} \left| p(r) 
-  \rho_c \left( \sqrt{1-2\hat m(\rho_c, r) /r} - K_c \over 3 K_c -  \sqrt{1-2\hat m(\rho_c, r) /r}\right)\right|
 \geq 0.
\end{equation}
Explicitly:
\begin{equation}
{d\over dr} \left| p(r) 
-  \rho_s \left( \sqrt{1-{8\pi\over3} \rho_s r^2} - K_s \over 3 K_s -  \sqrt{1-{8\pi\over3} \rho_s r^2}\right)\right|
 \leq 0;
\end{equation}
and
\begin{equation}
{d\over dr} \left| p(r) 
-  \rho_c \left( \sqrt{1-{8\pi\over3} \rho_c r^2} - K_c \over 3 K_c -  \sqrt{1-{8\pi\over3} \rho_c r^2}\right)\right|
 \geq 0.
\end{equation}
One is still free to choose the constants $K_c$ and $K_s$ to maximum advantage, \\
which we shall do in the various corollaries developed below. \hfill\hfill $\Box$

\clearpage
{\bf Corollary 2a:} \\
Let the surface of the star be located at $r_s$ so that $p(r_s)=0$. \\
Choose $K_s =\sqrt{1-{8\pi\over3} \rho_s r_s^2}$ and $K_c =\sqrt{1-{8\pi\over3} \rho_c r_s^2}$. \\

Then
\begin{equation}
\label{E:hi2}
\hat p(\rho_s, K_s; r) =  \rho_s \left( \sqrt{1-{8\pi\over3} \rho_s r^2} - \sqrt{1-{8\pi\over3} \rho_s r_s^2}
\over 3 \sqrt{1-{8\pi\over3} \rho_s r_s^2}-  \sqrt{1-{8\pi\over3} \rho_s r^2}\right);
\end{equation}
with
\begin{equation}
\hat p(\rho_s, K_s; r_s) = 0; \qquad
\left.{d\hat p(\rho_s, K_s; r) \over dr}\right|_{r_s} = 
- {{4\pi\over 3} \rho_s^2 r_s^2\over r_s^2(1-{8\pi\over 3}\rho_s r_s^2)};
\end{equation}
and 
\begin{equation}
\label{E:low2}
\hat p(\rho_c, K_c; r) =  \rho_s \left( \sqrt{1-{8\pi\over3} \rho_c r^2} - \sqrt{1-{8\pi\over3} \rho_c r_s^2}
\over 3 \sqrt{1-{8\pi\over3} \rho_c r_s^2}-  \sqrt{1-{8\pi\over3} \rho_c r^2}\right);
\end{equation}
with
\begin{equation}
\hat p(\rho_c, K_c; r_s) = 0; \qquad
\left.{d\hat p(\rho_c, K_c; r) \over dr}\right|_{r_s} = 
- {{4\pi\over 3} \rho_c^2 r_s^2\over r_c^2(1-{8\pi\over 3} \rho_c r_s^2)}.
\end{equation}
Now note that because of our bounded positive density assumption we have
\begin{equation}
{{4\pi\over 3} \rho_c^2 r_s^2\over r_c^2(1-{8\pi\over 3} \rho_c r_s^2)} > 
{\rho_s m_s\over r_s^2 (1-2m_s/r_s)}
> {{4\pi\over 3} \rho_s^2 r_s^2\over r_s^2(1-{8\pi\over 3} \rho_s r_s^2)},
\end{equation}
thereby implying both
\begin{equation}
\hat p(\rho_c, K_c; r_s)=  p(r_s) =  \hat p(\rho_s, K_s; r_s) = 0,
\end{equation}
and
\begin{equation}
\left.{d\hat p(\rho_c, K_c; r) \over dr}\right|_{r_s} < 
\left.{dp(r) \over dr}\right|_{r_s}
< \left.{d\hat p(\rho_s, K_s; r) \over dr}\right|_{r_s}.
\end{equation}

Then integrating the differential inequality on $p(r)-\hat p(r)$ downwards from the surface $r_s$ towards the centre, taking account of the boundary conditions on  $\hat p(\rho_c, K_c; r)$,   $p(r)$, and $\hat p(\rho_s, K_s; r)$, and their derivatives, we see
{\small
\begin{eqnarray}
\label{E:high-low}
&&
\!\!\!\!\!\!\!\!
\rho_s \left( \sqrt{1-{8\pi\over3} \rho_s r^2} - \sqrt{1-{8\pi\over3} \rho_s r_s^2}
\over 3 \sqrt{1-{8\pi\over3} \rho_s r_s^2}-  \sqrt{1-{8\pi\over3} \rho_s r^2}\right)
\leq 
p(r) \leq  
\rho_c \left( \sqrt{1-{8\pi\over3} \rho_c r^2} - \sqrt{1-{8\pi\over3} \rho_c r_s^2}
\over 3 \sqrt{1-{8\pi\over3} \rho_c r_s^2}-  \sqrt{1-{8\pi\over3} \rho_c r^2}\right);
\nonumber\\
&& \!\!\!\!\!\!\!\!\!\!\!\!\!\!\!\!\!\!\!\!\!\!\!\!\!\!\!\!
\end{eqnarray}}
\vspace{-5pt}
with equality only at the surface $r=r_s$. 

\clearpage
{\bf Note 2a-1:}\\
Observe  that for small $\rho_c$ and $\rho_s$ we have
\begin{equation}
{2\pi\over3} \rho_s^2 [r_s^2-r^2] + \O(\rho_s^3) < p(r) < {2\pi\over3} \rho_c^2 [r_s^2-r^2] + \O(\rho_c^3).
\end{equation}
These low-density approximations are easily seen to be simple adaptations of the pressure profile for a Newtonian constant density sphere:
\begin{equation}
p(r) = {2\pi\over3} \rho_0^2 [r_s^2 -r^2].
\end{equation}
See for instance references~\cite{Chandrasekhar:1942} or~\cite[page 128]{Wald:1984}.

{\bf Note 2a-2:}\\
Observe that  these bounds (\ref{E:high-low}) are tightest at the surface of the star, where both upper and lower bounds converge on zero, and weakest at the centre of the star, where we have
\begin{equation}
\rho_s \left( 1 - \sqrt{1-{8\pi\over3} \rho_s r_s^2}
\over 3 \sqrt{1-{8\pi\over3} \rho_s r_s^2}-  1\right)
<p_c <  
\rho_c \left( 1 - \sqrt{1-{8\pi\over3} \rho_c r_s^2}
\over 3 \sqrt{1-{8\pi\over3} \rho_c r_s^2}-  1\right).
\end{equation}

{\bf Note 2a-3:}\\
Observe that the central pressure definitely diverges as
\begin{equation}
3 \sqrt{1-{8\pi\over3} \rho_c r_s^2}-  1 \;\;\to\;\; 0;
\qquad \implies \qquad
{8\pi\over3} \rho_c r_s^2 \;\;\to\;\; {8\over 9}. 
\end{equation}
This is closely related to, but slightly weaker than, the famous Buchdahl--Bondi bound~\cite{Buchdahl:1959,Bondi:1964}.

{\bf Note 2a-4:}\\
Observe  that for small $\rho_c$ and $\rho_s$ the central pressure bounds reduce to
\begin{equation}
{2\pi\over3} \rho_s^2 r_s^2 + \O(\rho_s^3) < p_c < {2\pi\over3} \rho_c^2 r_s^2 + \O(\rho_c^3).
\end{equation}
These are easily seen to be simple adaptations of the known result for the central pressure of a Newtonian constant density sphere:
\begin{equation}
p_c = {2\pi\over3} \rho_0^2 r_s^2.
\end{equation}
See for instance~\cite{Chandrasekhar:1942} or~\cite[page 128]{Wald:1984}.

\clearpage
{\bf Corollary 2b:} \\
Let the central pressure be $p(0)=p_c$. \\
Choose $K_s =  (\rho_s+p_c)/(\rho_s+3p_c)   $ and $K_c = (\rho_c+p_c)/(\rho_c+3p_c) $. \\
Then
\begin{equation}
\hat p(\rho_s,K_s; r) =  \rho_s \left( [\rho_s+3 p_c] \sqrt{1-{8\pi\over3} \rho_s r^2} - [\rho_s + p_c]
\over 3 [\rho_s + p_c] -  [\rho_s+3 p_c] \sqrt{1-{8\pi\over3} \rho_s r^2}\right);
\end{equation}
with
\begin{equation}
\left.\hat p(\rho_s,K_s; r)\right|_{r=0}= p_c;
\qquad \hbox{and} \qquad
\left. {d^2\hat p(\rho_s,K_s; r)\over dr^2}\right|_{r=0} = - {4\pi\over 3} (\rho_s+p_c)(\rho_s+2p_c);
\end{equation}
and
\begin{equation}
\hat p(\rho_c,K_c; r) = \rho_c \left( [\rho_c+3 p_c] \sqrt{1-{8\pi\over3} \rho_c r^2} - [\rho_c+ p_c] 
\over 3 [\rho_c+ p_c] -  [\rho_c+3 p_c] \sqrt{1-{8\pi\over3} \rho_c r^2}\right);
\end{equation}
with
\begin{equation}
\left.\hat p(\rho_c,K_c; r)\right|_{r=0}= p_c;
\qquad \hbox{and} \qquad
\left. {d^2\hat p(\rho_c,K_c; r)\over dr^2}\right|_{r=0} =   - {4\pi\over 3} (\rho_c+p_c)(\rho_c+2p_c).
\end{equation}
We note that
\begin{equation}
\left. {d^2\hat p(\rho_c,K_c; r)\over dr^2}\right|_0  = \left. {d^2 p(r)\over dr^2}\right|_0 < \left. {d^2\hat p(\rho_s,K_s; r)\over dr^2}\right|_0 .
\end{equation}
Then, observing that for realistic stars $\rho(r) < \rho_c$ once one moves away from the centre,  a somewhat more delicate analysis (see the Appendix for the somewhat messy details) leads to the improved bounds
\begin{equation}
\left. {d^2\hat p(\rho_c,K_c; r)\over dr^2}\right|_0  < \left. {d^2 p(r)\over dr^2}\right|_0 < \left. {d^2\hat p(\rho_s,K_s; r)\over dr^2}\right|_0 .
\end{equation}
In view of this, integrating the differential inequality upwards from the centre towards the surface, we see
{\small
\begin{eqnarray}
&&
 \rho_c \left( [\rho_c+3 p_c] \sqrt{1-{8\pi\over3} \rho_c r^2} - [\rho_c+ p_c] 
\over 3 [\rho_c+ p_c] -  [\rho_c+3 p_c] \sqrt{1-{8\pi\over3} \rho_c r^2}\right)
\leq p(r) \leq
\rho_s \left( [\rho_s+3 p_c] \sqrt{1-{8\pi\over3} \rho_s r^2} - [\rho_s + p_c]
\over 3 [\rho_s + p_c] -  [\rho_s+3 p_c] \sqrt{1-{8\pi\over3} \rho_s r^2}\right); \nonumber\\
&&
\end{eqnarray}}
with equality only at the centre $r=0$.

\clearpage
{\bf Note 2b-1:}\\
Observe  that for small $\rho_c$ and $\rho_s$ we have
\begin{equation}
{p_c[1-{8\pi\over 3}\rho_c r^2 + \O(\rho_c^2)]\over1+2\pi p_c r^2}
\leq p(r) \leq 
{p_c[1-{8\pi\over 3}\rho_s r^2 + \O(\rho_s^2)]\over1+2\pi p_c r^2}
\end{equation}
As $\rho_c\to\rho_s\to 0$ this reproduces part of the discussion of Theorem 1. \\
Specifically, one regains Schwarzschild's zero density star~\cite{revisiting}.

{\bf Note 2b-2:}\\
Note these bounds are now tightest at the centre of the star, where both upper and lower bounds converge on $p_c$, and weakest at the surface of the star.\\
Specifically, at the surface
{\small
\begin{eqnarray}
&&
 \rho_c \left( [\rho_c+3 p_c] \sqrt{1-{8\pi\over3} \rho_c r_s^2} - [\rho_c+ p_c] 
\over 3 [\rho_c+ p_c] -  [\rho_c+3 p_c] \sqrt{1-{8\pi\over3} \rho_c r_s^2}\right)
> 0 >
\rho_s \left( [\rho_s+3 p_c] \sqrt{1-{8\pi\over3} \rho_s r_s^2} - [\rho_s + p_c]
\over 3 [\rho_s + p_c] -  [\rho_s+3 p_c] \sqrt{1-{8\pi\over3} \rho_s r_s^2}\right).\nonumber\\
&&
\end{eqnarray}}
Thence (since the denominators are guaranteed to be positive)
\begin{equation}
[\rho_c+3 p_c] \sqrt{1-{8\pi\over3} \rho_c r_s^2} - [\rho_c+ p_c]
> 0 >
[\rho_s+3 p_c] \sqrt{1-{8\pi\over3} \rho_s r_s^2} - [\rho_s + p_c],
\end{equation}
whence
\begin{equation}
{3\over2\pi} \; {p_c (\rho_c+2p_c)\over \rho_c(\rho_c+3p_c)^2 }
 < r_s^2 < 
 {3\over2\pi} \; {p_c (\rho_s+2p_c)\over \rho_s(\rho_s+3p_c)^2 }.
\end{equation}
In particular this implies the weaker but considerably simpler bounds
\begin{equation}
{1\over3\pi\rho_c} 
 < r_s^2 < 
 {1\over3\pi\rho_s}.
\end{equation}

{\bf Corollary 2c:} \\
Suppose now that we know the pressure $p_*=p(r_*)$ at some point $r_*\in(0,r_s)$ inside the star.
Choose 
\begin{equation}
K_s =  \sqrt{1-{8\pi\over3} \rho_s r_*^2}\;\;{(\rho_s+p_*)\over(\rho_s+3p_*)};
\qquad
K_c = 
\sqrt{1-{8\pi\over3} \rho_c r_*^2}\;\;{(\rho_c+p_*)\over(\rho_c+3p_*)}.
\end{equation}
Then
\begin{equation}
\hat p(\rho_s,K_s; r) =  \rho_s \left( [\rho_s+3 p_*] \sqrt{1-{8\pi\over3} \rho_s r^2} - [\rho_s + p_*]\sqrt{1-{8\pi\over3} \rho_s r_*^2}
\over 3 [\rho_s + p_*]\sqrt{1-{8\pi\over3} \rho_s r_*^2} -  [\rho_s+3 p_*] \sqrt{1-{8\pi\over3} \rho_s r^2}\right);
\end{equation}
with $\hat p(\rho_s,K_s; r_*) = p_*$  and
\begin{equation}
\label{E:slope-star-to-surface}
\left.{d \hat p(\rho_s,K_s; r) \over dr}\right|_{r_*} =
 -{4\pi\over 3} {(\rho_s+p_*)(\rho_s+3p_*) r_* \over 1- {8\pi\over3} \rho_s r_*^2}.
\end{equation}

Similarly
\begin{equation}
\hat p(\rho_c,K_c; r) = \rho_c \left( [\rho_c+3 p_*] \sqrt{1-{8\pi\over3} \rho_c r^2} - [\rho_c+ p_*] \sqrt{1-{8\pi\over3} \rho_c r_*^2}
\over 3 [\rho_c+ p_*] \sqrt{1-{8\pi\over3} \rho_c r_*^2}-  [\rho_c+3 p_*] \sqrt{1-{8\pi\over3} \rho_c r^2}\right);
\end{equation}
with $\hat p(\rho_c,K_c; r_*) = p_*$ and
\begin{equation}
\label{E:slope-center-to-star}
\left.{d \hat p(\rho_c,K_c; r) \over dr}\right|_{r_*} =
 -{4\pi\over 3} {(\rho_c+p_*)(\rho_c+3p_*) r_* \over 1- {8\pi\over3} \rho_c r_*^2}.
\end{equation}
Comparing equations (\ref{E:slope-center-to-star}), (\ref{E:slope-at-star}),  and  (\ref{E:slope-star-to-surface}) we note that under the assumption of bounded positive density we have 
\begin{equation}
\left.{d \hat p(\rho_c,K_c; r) \over dr}\right|_{r_*}  < \left.{d p(r) \over dr}\right|_{r_*} < \left.{d \hat p(\rho_s,K_s; r) \over dr}\right|_{r_*}.
\end{equation}

Then, using the monotonicity of $|p(r)-\hat p(r)|$ as per Theorem 2,   for $r\in(r_*,r_s)$ we have
\begin{equation}
\label{E:C2c-eq1}
p(r) <  \rho_s \left( [\rho_s+3 p_*] \sqrt{1-{8\pi\over3} \rho_s r^2} - [\rho_s + p_*]\sqrt{1-{8\pi\over3} \rho_s r_*^2}
\over 3 [\rho_s + p_*]\sqrt{1-{8\pi\over3} \rho_s r_*^2} -  [\rho_s+3 p_*] \sqrt{1-{8\pi\over3} \rho_s r^2}\right);
\end{equation}
and
\begin{equation}
\label{E:C2c-eq2}
p(r) >  \rho_c \left( [\rho_c+3 p_*] \sqrt{1-{8\pi\over3} \rho_c r^2} - [\rho_c+ p_*] \sqrt{1-{8\pi\over3} \rho_c r_*^2}
\over 3 [\rho_c+ p_*] \sqrt{1-{8\pi\over3} \rho_c r_*^2}-  [\rho_c+3 p_*] \sqrt{1-{8\pi\over3} \rho_c r^2}\right).
\end{equation}
In counterpoint, for $r\in(0,r_*)$ the direction of the inequalities is reversed
\begin{equation}
\label{E:C2c-eq3}
p(r) >  \rho_s \left( [\rho_s+3 p_*] \sqrt{1-{8\pi\over3} \rho_s r^2} - [\rho_s + p_*]\sqrt{1-{8\pi\over3} \rho_s r_*^2}
\over 3 [\rho_s + p_*]\sqrt{1-{8\pi\over3} \rho_s r_*^2} -  [\rho_s+3 p_*] \sqrt{1-{8\pi\over3} \rho_s r^2}\right);
\end{equation}
and
\begin{equation}
\label{E:C2c-eq4}
p(r) <  \rho_c \left( [\rho_c+3 p_*] \sqrt{1-{8\pi\over3} \rho_c r^2} - [\rho_c+ p_*] \sqrt{1-{8\pi\over3} \rho_c r_*^2}
\over 3 [\rho_c+ p_*] \sqrt{1-{8\pi\over3} \rho_c r_*^2}-  [\rho_c+3 p_*] \sqrt{1-{8\pi\over3} \rho_c r^2}\right).
\end{equation}

{\bf Notes 2c-1:}\\
As $r\to r_*$, from above or below, both upper and lower bounds converge on $p_*$.\\
As $r_*\to0$, since $p_*\to p_c$,  one recovers the bounds of Corollary 2b.\\
As $r_*\to r_s$, since $p_*\to 0$, one recovers the bounds of Corollary 2a.\\
For small densities one again recovers part of Theorem 1
\begin{equation}
p(r) < {p_*\over 1+ 2\pi p_* [r^2-r_*^2] + \O(\rho_s) }\qquad (r \in (r_*,r_s) );
\end{equation}
\begin{equation}
p(r) > {p_*\over 1+ 2\pi p_* [r^2-r_*^2] + \O(\rho_c) }\qquad (r \in (r_*,r_s) );
\end{equation}
and
\begin{equation}
p(r) > {p_*\over 1+ 2\pi p_* [r^2-r_*^2]- \O(\rho_s)} \qquad (r \in (0,r_*) );
\end{equation}
\begin{equation}
p(r) < {p_*\over 1+ 2\pi p_* [r^2-r_*^2] - \O(\rho_c) }\qquad (r \in (0,r_*) );
\end{equation}
where one can arrange $\O(\rho_c) > \O(\rho_s) > 0$.
Specifically $\rho_s\to 0$ recovers equations (\ref{E:T1a}) and (\ref{E:T1b}), while $\rho_c\to 0$ reproduces Schwarzschild's zero density star~\cite{revisiting}. 

{\bf Note 2c-2:} \\
In Corollary 2c we now take $r\to0$ keeping both $r_*$ and $p_*$ fixed. \\
Then we can bound the central pressure as
\begin{equation}
p_c >  \rho_s \left( [\rho_s+3 p_*] - [\rho_s + p_*]\sqrt{1-{8\pi\over3} \rho_s r_*^2}
\over 3 [\rho_s + p_*]\sqrt{1-{8\pi\over3} \rho_s r_*^2} -  [\rho_s+3 p_*] \right)
\end{equation}
and
\begin{equation}
p_c <  \rho_c \left( [\rho_c+3 p_*]  - [\rho_c+ p_*] \sqrt{1-{8\pi\over3} \rho_c r_*^2}
\over 3 [\rho_c+ p_*] \sqrt{1-{8\pi\over3} \rho_c r_*^2}-  [\rho_c+3 p_*] \right)
\end{equation}

For small $\rho_s$ or $\rho_c$ one finds
{\small
\begin{equation}
{p_*\over1-2\pi p_* r_*^2} \left( 1 + {{8\pi\over3}\rho_c r_*^2\over1-2\pi p_* r_*^2} +\O(\rho_c^2)\right)
<p_c < 
{p_*\over1-2\pi p_* r_*^2} \left( 1 + {{8\pi\over3}\rho_c r_*^2\over1-2\pi p_* r_*^2} +\O(\rho_c^2)\right).
\end{equation}
}
Specifically, for $\rho_c\to \rho_s\to 0$ one finds
\begin{equation}
p_c \to
{p_*\over1-2\pi p_* r_*^2},
\end{equation}
reproducing part of the discussion of Theorem 1.


{\bf Comment on Theorem 2:} \\
Note that none of the bounds developed from Theorem 2 or its corollaries have made any use of the quantity $\rho_*=\rho(r_*)$.
Ultimately this is because of the extreme weakness of our key input assumption --- \emph{boundedness of $\rho(r)$ but without any monotonicity demand} --- so even if we know  $\rho_*=\rho(r_*)$ under the current hypotheses this gives us no information as to how $\rho(r)$ is behaving in the vicinity of $r_*$. In view of this we shall now explore what happens once one adds extra input assumptions in the form of a a monotonicity assumption.

\clearpage
\section{Theorem 3 (bounded monotone positive density)}

Let us now demand that the density be both bounded and monotone decreasing as one moves outwards.
Specifically:
\begin{equation}
 \rho_c > \rho(r) > \rho_s >0, \qquad \hbox{and}\qquad  {d\rho(r)\over dr} < 0,\qquad \hbox{for} \qquad r\in(0,r_s).
\end{equation}
Suppose now that we know the pressure $p_*=p(r_*)$ and density $\rho_*=\rho(r_*)$ at some point $r_*\in(0,r_s)$ inside the star. We can now deduce both
\begin{equation}
\rho_c > \rho(r) > \rho_* \qquad\hbox{for}\qquad r\in(0,r_*);
\end{equation}
and
\begin{equation}
\rho_* > \rho(r) > \rho_s \qquad\hbox{for}\qquad r\in(r_*,r_s).
\end{equation}
This is more information than we previously had, because now we can make some statements about $\rho_*=\rho(r_*)$.

Under these circumstances the \emph{two} key differential inequalities (\ref{E:diff-ineq1})--(\ref{E:diff-ineq2}) in the discussion leading up to Theorem 2 can be piecewise refined to yield \emph{four} differential inequalities
\begin{equation}
{dp(r)\over dr} <
- {[\rho_* +p(r)]\; [\hat m(\rho_*, r) +4\pi p(r) r^3]\over 
r^2(1-2\hat m(\rho_*, r) /r)}; \qquad r\in (0,r_*);
\end{equation}
\begin{equation}
{dp(r)\over dr} <
- {[\rho_s +p(r)]\; [\hat m(\rho_s, r) +4\pi p(r) r^3]\over 
r^2(1-2\hat m(\rho_s, r) /r)}; \qquad r\in (r_*,r_s);
\end{equation}
and 
\begin{equation}
{dp(r)\over dr} > 
- {[\rho_c +p(r)]\; [\hat m(\rho_c, r) +4\pi p(r) r^3]\over 
r^2(1-2\hat m(\rho_c, r) /r)};  \qquad r\in (0,r_*);
\end{equation}
\begin{equation}
{dp(r)\over dr} > 
- {[\rho_* +p(r)]\; [\hat m(\rho_*, r) +4\pi p(r) r^3]\over 
r^2(1-2\hat m(\rho_*, r) /r)}; \qquad r\in (r_*,r_s).
\end{equation}

It is still useful to consider the quantity
\begin{equation}
\hat p(\rho_0, K_0, r) = \rho_0 \left( \sqrt{1-{8\pi\over3} \rho_0 r^2} - K_0 \over 3 K_0 -  \sqrt{1-{8\pi\over3} \rho_0 r^2}\right),
\end{equation}
which we can now use to develop Theorem 3 below.

\clearpage
{\bf Theorem 3:}\\
For a bounded and monotone density profile (as characterized above) we have:
\begin{equation}
{d\over dr} \left| p(r) 
-  \rho_* \left( \sqrt{1-{8\pi\over3} \rho_* r^2} - K_* \over 3 K_* -  \sqrt{1-{8\pi\over3} \rho_* r^2}\right)\right|
 <0; \qquad r \in(0,r_*);
\end{equation}
\begin{equation}
{d\over dr} \left| p(r) 
-  \rho_s \left( \sqrt{1-{8\pi\over3} \rho_s r^2} - K_s \over 3 K_s -  \sqrt{1-{8\pi\over3} \rho_s r^2}\right)\right|
 <0; \qquad r \in(r_*, r_s);
\end{equation}
and
\begin{equation}
{d\over dr} \left| p(r) 
-  \rho_c \left( \sqrt{1-{8\pi\over3} \rho_c r^2} - K_c \over 3 K_c -  \sqrt{1-{8\pi\over3} \rho_c r^2}\right)\right|
 >0; \qquad r \in(0,r_*);
\end{equation}
\begin{equation}
{d\over dr} \left| p(r) 
-  \rho_* \left( \sqrt{1-{8\pi\over3} \rho_* r^2} - K_* \over 3 K_* -  \sqrt{1-{8\pi\over3} \rho_* r^2}\right)\right|
 >0; \qquad r \in(r_*,r_s);
\end{equation}
One is still free to choose the constants $K_*$, $K_c$, and $K_s$ to maximum advantage. \hfill\hfill $\Box$

{\bf Corollary 3a:}\\
The analogue of Corollary 2a would correspond  to setting $r_*\to r_s$ and $\rho_*\to \rho_s$. \\
This would merely reproduce Corollary 2a providing no extra information.

{\bf Corollary 3b:} \\
The analogue of Corollary 2b would correspond  to setting $r_*\to 0$ and $\rho_*\to \rho_c$. \\
This would merely reproduce Corollary 2b providing no extra information.

{\bf Corollary 3c:}\\
The analogue of Corollary 2c is more interesting.

Choose 
\begin{equation}
K_s =  K_c = \sqrt{1-{8\pi\over3} \rho_* r_*^2}\;\;{(\rho_*+p_*)\over(\rho_*+3p_*)}.
\end{equation}
Then for $r\in(r_*,r_s)$ we have
\begin{equation}
p(r) <  \rho_* \left( [\rho_*+3 p_*] \sqrt{1-{8\pi\over3} \rho_* r^2} - [\rho_*+ p_*] \sqrt{1-{8\pi\over3} \rho_* r_*^2}
\over 3 [\rho_*+ p_*] \sqrt{1-{8\pi\over3} \rho_* r_s^2}-  [\rho_*+3 p_*] \sqrt{1-{8\pi\over3} \rho_* r^2}\right).
\end{equation}
(That is, we can now replace $\rho_s\to\rho_*$ in equation (\ref{E:C2c-eq1}).)\\
In counterpoint, for $r\in(0,r_*)$ the direction of the inequality is reversed
\begin{equation}
p(r) >  \rho_* \left( [\rho_*+3 p_*] \sqrt{1-{8\pi\over3} \rho_* r^2} - [\rho_*+ p_*] \sqrt{1-{8\pi\over3} \rho_* r_*^2}
\over 3 [\rho_*+ p_*] \sqrt{1-{8\pi\over3} \rho_* r_s^2}-  [\rho_*+3 p_*] \sqrt{1-{8\pi\over3} \rho_* r^2}\right).
\end{equation}
(That is, we can now replace $\rho_s\to\rho_*$ in equation (\ref{E:C2c-eq3}).)\\

{\bf Notes 3c-1:}\\
As $r\to r_*$, from either above or below,  upper and lower bounds converge on $p_*$.\\
As $r_*\to0$, since $p_*\to p_c$ and $\rho_*\to \rho_c$,  one recovers the bounds of Corollary 2b.\\
As $r_*\to r_s$, since $p_*\to 0$ and $\rho_*\to\rho_s$, one recovers the bounds of Corollary 2a.\\
But for $r_*\in(0,r_s)$ the current bounds provide genuinely more information. \\
For small density $\rho_*$ one again recovers results of Theorem 1:
\begin{equation}
p(r) < {p_*\over 1+ 2\pi p_* [r^2-r_*^2] + \O(\rho_*) }\qquad (r \in (r_*,r_s) );
\end{equation}
and
\begin{equation}
p(r) > {p_*\over 1+ 2\pi p_* [r^2-r_*^2]- \O(\rho_*)} \qquad (r \in (0,r_*) );
\end{equation}
where one can arrange $ \O(\rho_*) > 0$.

{\bf Note 3c-2:}\\
In Corollary 3c take $r\to0$ keeping both $r_*$ and $p_*$ fixed. \\
Then we can bound the central pressure as
\begin{equation}
p_c >  \rho_* \left( [\rho_*+3 p_*]  - [\rho_*+ p_*] \sqrt{1-{8\pi\over3} \rho_* r_*^2}
\over 3 [\rho_*+ p_*] \sqrt{1-{8\pi\over3} \rho_* r_s^2}-  [\rho_*+3 p_*] \right).
\end{equation}
For small $\rho_*$  one finds
\begin{equation}
p_c > {p_*\over1-2\pi p_* r_*^2} \left( 1 + {{8\pi\over3}\rho_* r_*^2\over1-2\pi p_* r_*^2} +\O(\rho_*^2)\right).
\end{equation}
Specifically, for $\rho_*\to 0$ one again finds
\begin{equation}
p_c \to
{p_*\over1-2\pi p_* r_*^2},
\end{equation}
reproducing part of the discussion of Theorem 1.

{\bf Comment:}\\
As is to be expected, we have been able to say more about the pressure profile by adding more input assumptions. 

\enlargethispage{50pt}

\section{Theorem 4 (bounded monotone positive average density)}
 
 Can we say anything more restrictive when dealing with the average density
 \begin{equation}
\bar\rho(r) = {m(r)\over{4\pi\over 3} r^3} = 
{\int_0^r 4\pi \tilde r^2 \rho(\tilde r) d\tilde r \over\int_0^r 4\pi \tilde r^2 d\tilde r }\;\;?
\end{equation}
 Start (as per Theorem 2) by assuming only minimalist boundedness conditions on the density itself
\begin{equation}
 \rho_c > \rho(r) > \rho_s \qquad \hbox{for} \qquad r\in(0,r_s).
\end{equation}
Then by definition we have:
\begin{equation}
m(r) = \int_0^r 4\pi \rho(\tilde r) \tilde r^2 d\tilde r = {4\pi\over 3} \bar\rho(r) r^3 .
\end{equation}
Thence we separately have  both
\begin{equation}
\rho_c > \bar\rho(r) > \rho_s
\qquad \hbox{and also} \qquad
\rho_c >  \rho(r) > \rho_s.
\end{equation}
So boundedness of the density implies boundedness of the average density.

But based solely on this information alone one cannot (yet) decide whether
\begin{equation}
\bar\rho(r) < \rho(r), \qquad \hbox{or} \qquad \bar\rho(r) > \rho(r), \qquad \hbox{or might the inequalities oscillate?}
\end{equation}
Based solely on this information alone we can only deduce
\begin{equation}
\hat m(\rho_c,r) = {4\pi\over 3} \rho_c r^3  \geq m(r)  =  {4\pi\over 3} \bar\rho(r) r^3 
\geq  {4\pi\over 3} \rho_s r^3 = \hat m(\rho_s,r),
\end{equation}
but this would merely reproduce the discussion of Theorem 2.
To make progress, one needs at least one extra piece of information. 
There are at least three ways of getting the extra information we need:
\begin{itemize}
\item One could simply postulate $\bar\rho(r) > \bar\rho_s$ in which case 
\begin{equation}
m(r) > {4\pi\over 3} \bar\rho(r) r^3 
> {4\pi\over 3} \bar\rho_s r^3 = \hat m(\bar \rho_s,r).
\end{equation}
\item
More restrictively, one could simply postulate that volume averaged density $\bar\rho(r)$ be monotone decreasing, in which case $\bar\rho(r) > \bar\rho_s$, and (as above) it follows that $m(r) \geq  \hat m(\bar \rho_s,r)$
\item Even more restrictively, we can easily show that postulating monotonicity of the density $\rho(r)$,  (as per Theorem 3), automatically implies $\bar\rho(r) > \rho(r)$. \\
(Note that averaging a monotone function yields a monotone average.) 

In particular
\begin{equation}
{4\pi\over 3} \bar\rho_c r^3  > m(r) = {4\pi\over 3} \bar\rho(r) r^3 > {4\pi\over 3} \rho(r) r^3.
\end{equation}
In addition, evaluating the derivative
\begin{equation}
{dm(r)\over dr} = 4\pi \rho(r) r^2 = 4\pi \bar\rho(r) r^2 + {4\pi\over 3} {d \bar\rho(r)\over dr}  r^3,
\end{equation}
so
\begin{equation}
{d \bar\rho(r)\over dr}  = 3 \; {\rho(r) - \bar\rho(r)\over r} < 0,
\end{equation}
implying monotonicity of $\bar\rho(r)$.

So under these conditions we have
\begin{equation}
\rho_c > \bar\rho(r)>\rho(r)>\rho_s \quad \hbox{and} \quad \bar\rho'(r)<0 \quad\hbox{and}\quad  \rho'(r)<0.
\end{equation}

\end{itemize}
Under any of these three options we certainly have $m(r) \geq \hat m(\bar\rho_s,r)$, with equality only in the case of Schwarzschild's constant density star ($\rho(r) = \bar\rho_s = \rho_s$).

To apply these considerations to the TOV, first we observe
\begin{equation}
{[\rho(r)+p(r)]\,[m(r)+4\pi p(r) r^3]\over r^2[1-2m(r)/r]} > 
{[\rho_s +p(r)]\, [\hat m(r,\bar\rho_s ) + 4\pi p(r) r^3]\over r^2 [1-2\hat m(r,\bar\rho_s)/r]},
\end{equation}
and similarly
\begin{equation}
{[\rho(r)+p(r)]\,[m(r)+4\pi p(r) r^3]\over r^2[1-2m(r)/r]} < 
{([\rho_c +p(r)]\,[\hat m(r,\rho_c ) + 4\pi p(r) r^3]\over r^2 [1-2\hat m(r,\rho_c)/r]}.
\end{equation}
The second of these bounds (depending on $\rho_c$) is uninteresting in that it merely reproduces the lower bound 
used in Theorem 3. The first of these bounds, depending on \emph{both} $\rho_s$ and $\bar\rho_s$,  is more interesting.

This leads to the ``interesting'' differential inequality
\begin{equation}
{dp(r)\over d r} <
-{[\rho_s +p(r)]\,[\hat m(r,\bar\rho_s ) + 4\pi p(r) r^3]\over r^2 [1-2\hat m(r,\bar\rho_s)/r]}.
\end{equation}
Observe
\begin{equation}
-{[\rho_s +p(r)]\,[\hat m(r,\bar\rho_s ) + 4\pi p(r) r^3]\over r^2 [1-2\hat m(r,\bar\rho_s)/r]}
=
-{4\pi\over3} {[\rho_s +p(r)]\,[\bar\rho_s + 3p(r)] r
\over 
1-{8\pi\over3}\bar\rho_s r^2}.
\end{equation}
So the relevant differential inequality can be rewritten as 
\begin{equation}
{dp(r)\over d r} <
-{{4\pi\over 3} [\rho_s +p(r)]\,[\bar\rho_s + 3p(r)] r
\over 
1-{8\pi\over3}\bar\rho_s r^2}.
\end{equation}

Since now we can have $\bar\rho_s\neq \rho_s$ this is \emph{not} just the differential inequality relevant to any constant density star~\cite{Wald:1984, Schwarzschild-star, Schwarzschild-star-translation}, but it is still mathematically simple enough to permit (with some effort) successful analysis. 
(Physics warning: This differential inequality is now just a mathematical inequality --- it does not directly correspond to any physical model for a star.)
Mathematically it is now easy to see that the quantity
\begin{equation}
\hat p(\rho_s,\bar\rho_s,K; r) = 
 {\rho_s  \left(1-{8\pi\over 3} \bar\rho_s r^2\right)^{{3\rho_s\over  4 \bar\rho_s}-{1\over4}} - K \bar\rho_s
\over
3K - \left(1-{8\pi\over 3} \bar\rho_s r^2\right)^{{3\rho_s\over  4 \bar\rho_s}-{1\over4}}}
\end{equation}
or equivalently
\begin{equation}
\hat p(\rho_s,\bar\rho_s,K; r) = 
 {\rho_s  - K \bar\rho_s \left(1-{8\pi\over 3} \bar\rho_s r^2\right)^{{1\over4}-{3\rho_s\over  4 \bar\rho_s}  }
\over
3K \left(1-{8\pi\over 3} \bar\rho_s r^2\right)^{{1\over4}-{3\rho_s\over  4 \bar\rho_s}}-1}
\end{equation}

saturates this differential inequality for arbitrary $K$. 
Note the non-trivial exponent appearing herein
\begin{equation}
{{3\rho_s\over  4 \bar\rho_s}-{1\over4}} \;\;\to\;\;  {1\over2}
\qquad \hbox{as} \qquad
\bar\rho_s \to \rho_s.
\end{equation}
(Which would then reproduce the discussion of Theorem 3 above.)
Since under the current assumptions $\rho_s\in[0,\bar\rho_s]$ it is sometimes useful to write $\rho_s = w \bar\rho_s$ with $w\in[0,1]$. The non-trivial exponent then can be written as $(3w-1)/4$ which is sometimes more tractable. 

That is
\begin{equation}
\hat p(\rho_s,\bar\rho_s,K; r) = 
\bar\rho_s\;\;
 {w  \left(1-{8\pi\over 3} \bar\rho_s r^2\right)^{{3w-1\over  4}} - K 
\over
3K - \left(1-{8\pi\over 3} \bar\rho_s r^2\right)^{{3w-1\over  4}}},
\end{equation}
or equivalently
\begin{equation}
\hat p(\rho_s,\bar\rho_s,K; r) = 
\bar\rho_s\;\;
 {w   - K \left(1-{8\pi\over 3} \bar\rho_s r^2\right)^{{1-3w\over  4}}
\over
3K \left(1-{8\pi\over 3} \bar\rho_s r^2\right)^{{1-3w\over  4}}-1 }.
\end{equation}

So to keep the comparison pressure profile $\hat p(\rho_s,\bar\rho_s,K; r)$ finite and positive at the origin $r=0$ one needs either $w > K >1/3$ or $w<K<1/3$. 
This already makes it clear that the case $w=1/3$ is exceptional, and that some care must be taken in choosing both the constant $K$ and parameter $w$. 

It is useful to note that
\begin{equation}
\hat p(\rho_s,\bar\rho_s,K; 0) = {\bar\rho_s (w-K)\over3K-1};
\qquad
\hat p''(\rho_s,\bar\rho_s,K; 0) =-{4\pi\over3} \bar\rho_s^2 K \left(3w-1\over3K-1\right)^2.
\end{equation}

 We now note
 \begin{equation}
{d\over dr} [p(r)-\hat p(r)] <
-{[\rho_s +p(r)]\,[\hat m(r,\bar\rho_s ) + 4\pi p(r) r^3]\over r^2 [1-2\hat m(r,\bar\rho_s)/r]}
+ 
{[\rho_s +\hat p(r)]\,[\hat m(r,\bar\rho_s ) + 4\pi \hat p(r) r^3]\over r^2 [1-2\hat m(r,\bar\rho_s)/r]}
\end{equation}
So as usual
\begin{equation}
{d\over dr} [p(r)-\hat p(r)] < - [p(r)-\hat p(r)]  \times\hbox{(something positive)},
\end{equation}
implying
\begin{equation}
{d\over dr} |p(r)-\hat p(r)| \leq 0.
\end{equation}
We now use this result to formulate theorem 4 below.

{\bf Theorem 4:}\\
\begin{equation}
{d\over dr} \left|p(r)- {\rho_s  \left(1-{8\pi\over 3} \bar\rho_s r^2\right)^{{3\rho_s\over  4 \bar\rho_s}-{1\over4}} - K \bar\rho_s
\over
3K - \left(1-{8\pi\over 3} \bar\rho_s r^2\right)^{{3\rho_s\over  4 \bar\rho_s}-{1\over4}}}\right| < 0
\qquad (r<r_s).
\end{equation}
Given any fixed parameter $w=\rho_s/\bar\rho_s$, we are still free to choose $K$ to maximum advantage.
\hfill\hfill $\Box$

{\bf Corollary 4a:}\\
Choose
\begin{equation}
K_s = {\rho_s\over \bar \rho_s} \left(1-{8\pi\over 3} \bar\rho_s r_s^2\right)^{{3\rho_s\over  4 \bar\rho_s}-{1\over4}}
\end{equation}
so that
\begin{equation}
\hat p(\rho_s,\bar\rho_s,  K_s, r) 
=
\rho_s \;
{ \left(1-{8\pi\over 3} \bar\rho_s r^2\right)^{{3\rho_s\over  4 \bar\rho_s}-{1\over4}} 
-
 \left(1-{8\pi\over 3} \bar\rho_s r_s^2\right)^{{3\rho_s\over  4 \bar\rho_s}-{1\over4}}
 \over
 3 (\rho_s/\bar\rho_s)\left(1-{8\pi\over 3} \bar\rho_s r_s^2\right)^{{3\rho_s\over  4 \bar\rho_s}-{1\over4}}
 -
 \left(1-{8\pi\over 3} \bar\rho_s r^2\right)^{{3\rho_s\over  4 \bar\rho_s}-{1\over4}}
 }
\end{equation}
and in particular $\hat p(\rho_s,\bar\rho_s,  K_s, r_s) =0$ and
\begin{equation}
\hat p'(\rho_s,\bar\rho_s,  K_s, r_s) 
= - {4\pi\over3} { r_s \rho_s \bar\rho_s\over 1 - (8\pi/3) \bar\rho_s r_s^2}
= -{ \rho_s m_s\over r_s^2(1-2m_s/r_s)}.
\end{equation}
Thus at the surface $r_s$  the model $\hat p(\rho_s,\bar\rho_s,  K_s,r)$ fits both the value and the slope of the actual pressure profile $p(r)$. 

In view of this the best we can at this stage say is that \emph{either}
\begin{equation}
p(r) \geq \hat p(\rho_s,\bar\rho_s,  K_s,r)
\qquad \hbox{\emph{or}}\qquad
p(r) \leq \hat p(\rho_s,\bar\rho_s,  K_s,r),
\end{equation}
with equality only at the surface $r=r_s$. (No crossings are allowed.)
But as argued below only one of these two options is actually viable. 
Specifically one has 
\begin{equation}
p(r) \geq \rho_s
{ \left(1-{8\pi\over 3} \bar\rho_s r^2\right)^{{3\rho_s\over  4 \bar\rho_s}-{1\over4}} 
-
 \left(1-{8\pi\over 3} \bar\rho_s r_s^2\right)^{{3\rho_s\over  4 \bar\rho_s}-{1\over4}}
 \over
 3 \left(1-{8\pi\over 3} \bar\rho_s r_s^2\right)^{{3\rho_s\over  4 \bar\rho_s}-{1\over4}}
 -
 \left(1-{8\pi\over 3} \bar\rho_s r^2\right)^{{3\rho_s\over  4 \bar\rho_s}-{1\over4}}
 }
\end{equation}
with equality only at the surface $r=r_s$.
Alternatively
\begin{equation}
p(r) \geq  w \bar\rho_s\;\;
{ \left(1-{8\pi\over 3} \bar\rho_s r^2\right)^{{3w-1\over  4}} 
-
 \left(1-{8\pi\over 3} \bar\rho_s r_s^2\right)^{{3w-1\over  4}}
 \over
 3 \left(1-{8\pi\over 3} \bar\rho_s r_s^2\right)^{{3w-1\over 4}}
 -
 \left(1-{8\pi\over 3} \bar\rho_s r^2\right)^{{3w-1\over 4}}
 }.
\end{equation}
To see that the other $\leq$ option is \emph{non-viable}, consider the limit
 $\rho_s \to \bar\rho_s$, in which case one approaches a Schwarzschild constant density star, and \emph{violates} half of Corollary 2a.
Furthermore as $\rho_s \to {1\over3} \bar\rho_s$ from above one finds the \emph{non-viable} bound $\hat p \leq 0$.\\
Finally for $\rho_s < {1\over3} \bar\rho_s$ one finds the  \emph{non-viable} bound $\hat p \leq \hbox{(something negative)}$.
That is,  the $\leq$ option is \emph{non-viable}, and we are forced to adopt the  $\geq$ option. 

{\bf Note 4a-1:}\\
As $\rho_s \to \bar\rho_s$ (for the $\geq$ option) one approaches a Schwarzschild constant density star, \\ and recovers half of Corollary 2a. \\
As $\rho_s \to {1\over3} \bar\rho_s$ from above (for the $\geq$ option) one finds the viable but relatively uninteresting bound $\hat p \geq 0$.\\
For $\rho_s < {1\over3} \bar\rho_s$ (for the $\geq$ option) one finds the viable but relatively uninteresting bound $\hat p \geq \hbox{(something negative)}$.

{\bf Note 4a-2:}\\
At the center of the star we see
\begin{equation}
p_c \geq \rho_s
{ 1
-
 \left(1-{8\pi\over 3} \bar\rho_s r_s^2\right)^{{3\rho_s\over  4 \bar\rho_s}-{1\over4}}
 \over
 3 \left(1-{8\pi\over 3} \bar\rho_s r_s^2\right)^{{3\rho_s\over  4 \bar\rho_s}-{1\over4}}
 -
 1
 }
=
\rho_s
{ 1
-
 \left(1-2m_s /r_s\right)^{{3\rho_s\over  4 \bar\rho_s}-{1\over4}}
 \over
 3 \left(1-2m_s/r_s\right)^{{3\rho_s\over  4 \bar\rho_s}-{1\over4}}
 -
 1
 }.
\end{equation}
Note that the central pressure is guaranteed to diverge when
\begin{equation}
{2m_s\over r_s} \geq 1 - \left(1\over 3\right)^{1/\left({3\rho_s\over  4 \bar\rho_s}-{1\over4}\right)}
= 1 - \left(1\over 3\right)^{4\over  3w-1}
\end{equation}
As $\bar\rho_s\to\rho_s$ this again approaches the famous Buchdahl--Bondi bound~\cite{Buchdahl:1959,Bondi:1964}. \\
For $\rho_s < \bar\rho_s$ this  yields a slight  weakening of the Buchdahl--Bondi bound.

{\bf Corollary 4b:}\\
Choose
\begin{equation}
K = {\rho_s + p_c\over \bar\rho_s+3p_c},
\end{equation}
so that
\begin{equation}
\hat p(\rho_s,\bar\rho_s,  K, r) 
=
\rho_s \;
{\left(1-{8\pi\over 3} \bar\rho_s r^2\right)^{{3\rho_s\over  4 \bar\rho_s}-{1\over4}} 
-
 {\rho_s + p_c\over \bar\rho_s+3p_c} \over
 3  {\rho_s + p_c\over \bar\rho_s+3p_c} -
  \left(1-{8\pi\over 3} \bar\rho_s r^2\right)^{{3\rho_s\over  4 \bar\rho_s}-{1\over4}}
 }
\end{equation}
and in particular $\hat p(\rho_s,\bar\rho_s,  K, 0) =p_c$, while
\begin{equation}
\hat p''(\rho_s,\bar\rho_s,  K, 0) = -{4\pi\over3} (\bar\rho_s+3p_c)(\rho_s+p_c) >  -{4\pi\over3} (\rho_c+3p_c)(\rho_c+p_c) = p''(0).
\end{equation}

Now insert this $K$ into the differential inequality to obtain the bound:
\begin{equation}
p(r) \leq
\rho_s {\left(1-{8\pi\over 3} \bar\rho_s r^2\right)^{{3\rho_s\over  4 \bar\rho_s}-{1\over4}} 
-
 {\rho_s + p_c\over \bar\rho_s+3p_c} \over
 3  {\rho_s + p_c\over \bar\rho_s+3p_c} -
  \left(1-{8\pi\over 3} \bar\rho_s r^2\right)^{{3\rho_s\over  4 \bar\rho_s}-{1\over4}}
 },
\end{equation}
with equality only at the centre.

{\bf Note 4b-1:} \\
As $\rho_s \to \bar\rho_s$ one recovers half of Corollary 2b.\\
As $\rho_s \to \bar\rho_s/3$ one obtains the trivial bound $p(r) \leq p_c$.\\
For $\rho_s \in (0,1/3)\bar\rho_s$ the present analysis does not seem to lead to anything particularly useful.

{\bf Note 4b-2:} \\
At the surface of the star
\begin{equation}
0 < 
 \rho_s {\left(1-{8\pi\over 3} \bar\rho_s r_s^2\right)^{{3\rho_s\over  4 \bar\rho_s}-{1\over4}} - {\rho_s+ p_c\over \bar\rho_s+3p_c} \over
 3 {\rho_s + p_c\over \bar\rho_s+3p_c} -
  \left(1-{8\pi\over 3} \bar\rho_s r_s^2\right)^{{3\rho_s\over  4 \bar\rho_s}-{1\over4}}
 }
\end{equation}
This implies
\begin{equation}
\left(1-{8\pi\over 3} \bar\rho_s r_s^2\right) < \left({\rho_s+ p_c\over \bar\rho_s+3p_c}\right)^{-{3\rho_s\over  4 \bar\rho_s}+{1\over4}}
\end{equation}
whence
\begin{equation}
{2m_s\over r_s} = {8\pi\over 3} \bar\rho_s r_s^2 
> 1- \left({\rho_s+ p_c\over \bar\rho_s+3p_c}\right)^{-{3\rho_s\over  4 \bar\rho_s}+{1\over4}}
\end{equation}
This is again related to (but distinct from) the Buchdahl--Bondi bound~\cite{Buchdahl:1959,Bondi:1964}.

\clearpage
{\bf Corollary 4c:} \\
Suppose now that we know the pressure $p_*=p(r_*)$ at some point $r_*\in(0,r_s)$ inside the star.
Choose
\begin{equation}
K = {\rho_s + p_*\over \bar\rho_s+3p_*} \left(1-{8\pi\over 3} \bar\rho_s r_*^2\right)^{{3\rho_s\over  4 \bar\rho_s}-{1\over4}} 
\end{equation}
so that
\begin{equation}
\hat p(\rho_s,\bar\rho_s,  K, r) 
=
\rho_s \; 
 {\left(
\left(1-{8\pi\over 3} \bar\rho_s r^2\right)^{{3\rho_s\over  4 \bar\rho_s}-{1\over4}} 
-
 {\rho_s + p_*\over \bar\rho_s+3p_*} 
 \left(1-{8\pi\over 3} \bar\rho_s r_*^2\right)^{{3\rho_s\over  4 \bar\rho_s}-{1\over4}} 
 \over
 3  {\rho_s + p_*\over \bar\rho_s+3p_*} 
  \left(1-{8\pi\over 3} \bar\rho_s r_*^2\right)^{{3\rho_s\over  4 \bar\rho_s}-{1\over4}}
 -
\left(1-{8\pi\over 3} \bar\rho_s r^2\right)^{{3\rho_s\over  4 \bar\rho_s}-{1\over4}}  
\right)}
\end{equation}
and in particular $\hat p(\rho_s,\bar\rho_s,  K, r_*) =p_*$ while
\begin{equation}
\left.{d \hat p(\rho_s,\bar\rho_s,  K, r) \over dr}\right|_{r_*} = 
- {4\pi\over 3} {r_* (\bar\rho_s+3p_*)(\rho_s+p_*)\over 1 - (8\pi/3) \bar\rho_s r_*^2}
>
- {4\pi\over 3} {r_* (\bar\rho_*+3p_*)(\rho_*+p_*)\over 1 - (8\pi/3) \bar\rho_* r_*^2}.
\end{equation}
Thence
\begin{equation}
\left.{d \hat p(\rho_s,\bar\rho_s,  K, r) \over dr}\right|_{r_*} 
>
\left.{d  p( r) \over dr}\right|_{r_*}.
\end{equation}

Then the differential inequality implies
\begin{equation}
p(r) 
>
\rho_s \; 
 {\left(
\left(1-{8\pi\over 3} \bar\rho_s r^2\right)^{{3\rho_s\over  4 \bar\rho_s}-{1\over4}} 
-
 {\rho_s + p_*\over \bar\rho_s+3p_*} 
 \left(1-{8\pi\over 3} \bar\rho_s r_*^2\right)^{{3\rho_s\over  4 \bar\rho_s}-{1\over4}} 
 \over
 3  {\rho_s + p_*\over \bar\rho_s+3p_*} 
  \left(1-{8\pi\over 3} \bar\rho_s r_*^2\right)^{{3\rho_s\over  4 \bar\rho_s}-{1\over4}}
 -
\left(1-{8\pi\over 3} \bar\rho_s r^2\right)^{{3\rho_s\over  4 \bar\rho_s}-{1\over4}}  
\right)};
\qquad
(r\in(0,r_*));
\end{equation}
and
\begin{equation}
p(r) 
<
\rho_s \; 
 {\left(
\left(1-{8\pi\over 3} \bar\rho_s r^2\right)^{{3\rho_s\over  4 \bar\rho_s}-{1\over4}} 
-
 {\rho_s + p_*\over \bar\rho_s+3p_*} 
 \left(1-{8\pi\over 3} \bar\rho_s r_*^2\right)^{{3\rho_s\over  4 \bar\rho_s}-{1\over4}} 
 \over
 3  {\rho_s + p_*\over \bar\rho_s+3p_*} 
  \left(1-{8\pi\over 3} \bar\rho_s r_*^2\right)^{{3\rho_s\over  4 \bar\rho_s}-{1\over4}}
 -
\left(1-{8\pi\over 3} \bar\rho_s r^2\right)^{{3\rho_s\over  4 \bar\rho_s}-{1\over4}}  
\right)};
\qquad
(r\in(r_*,r_s)).
\end{equation}

{\bf Note 4c-1:} \\
As $\rho_s \to \bar\rho_s$ one recovers half of Corollary 2c.\\
As $\rho_s \to \bar\rho_s/3$ one recovers the trivial bound $p(r) < \infty$.

\bigskip
{\bf Note 4c-2:} \\
At the surface of the star
\begin{equation}
0
<
\rho_s \; 
 {\left(
\left(1-{8\pi\over 3} \bar\rho_s r_s^2\right)^{{3\rho_s\over  4 \bar\rho_s}-{1\over4}} 
-
 {\rho_s + p_*\over \bar\rho_s+3p_*} 
 \left(1-{8\pi\over 3} \bar\rho_s r_*^2\right)^{{3\rho_s\over  4 \bar\rho_s}-{1\over4}} 
 \over
 3  {\rho_s + p_*\over \bar\rho_s+3p_*} 
  \left(1-{8\pi\over 3} \bar\rho_s r_*^2\right)^{{3\rho_s\over  4 \bar\rho_s}-{1\over4}}
 -
\left(1-{8\pi\over 3} \bar\rho_s r_s^2\right)^{{3\rho_s\over  4 \bar\rho_s}-{1\over4}}  
\right)},
\end{equation}
implying
\begin{equation}
(\bar\rho_s+3p_*) \left(1-{8\pi\over 3} \bar\rho_s r_s^2\right)^{{3\rho_s\over  4 \bar\rho_s}-{1\over4}} 
-
 (\rho_s + p_*)
 \left(1-{8\pi\over 3} \bar\rho_s r_*^2\right)^{{3\rho_s\over  4 \bar\rho_s}-{1\over4}} > 0,
\end{equation}
whence
\begin{equation}
p_* > {\rho_s \left(1-{8\pi\over 3} \bar\rho_s r_*^2\right)^{{3\rho_s\over  4 \bar\rho_s}-{1\over4}} 
- \bar\rho_s \left(1-{8\pi\over 3} \bar\rho_s r_s^2\right)^{{3\rho_s\over  4 \bar\rho_s}-{1\over4}}
\over
3 \left(1-{8\pi\over 3} \bar\rho_s r_s^2\right)^{{3\rho_s\over  4 \bar\rho_s}-{1\over4}}
-
\left(1-{8\pi\over 3} \bar\rho_s r_*^2\right)^{{3\rho_s\over  4 \bar\rho_s}-{1\over4}}
}.
\end{equation}
This is compatible with Corollary 4a.

{\bf Note 4c-3:} \\
If in Corollary 4c we now take $r\to0$ keeping both $r_*$ and $p_*$ fixed, then we obtain a constraint on the central pressure
\begin{equation}
p_c
>
\rho_s \; 
 {\left(
1 -
 {\rho_s + p_*\over \bar\rho_s+3p_*} 
 \left(1-{8\pi\over 3} \bar\rho_s r_*^2\right)^{{3\rho_s\over  4 \bar\rho_s}-{1\over4}} 
 \over
 3  {\rho_s + p_*\over \bar\rho_s+3p_*} 
  \left(1-{8\pi\over 3} \bar\rho_s r_*^2\right)^{{3\rho_s\over  4 \bar\rho_s}-{1\over4}}
 - 1
\right)}.
\end{equation}
Equivalently
\begin{equation}
p_c
>
w \bar\rho_s \; 
 {\left(
1 -
 {w\bar\rho_s + p_*\over \bar\rho_s+3p_*} 
 \left[1-(2m_s/r_s) (r_*/r_s)^2\right]^{{3w-1\over 4}} 
 \over
 3  {w \bar\rho_s + p_*\over \bar\rho_s+3p_*} 
  \left[1-(2m_s/r_s) (r_*/r_s)^2\right]^{{3w-1\over  4}}
 - 1
\right)}.
\end{equation}
Though somewhat complex this does provide a bound on the central pressure $p_c$ in terms of surface data $(m_s,r_s,w)$, with $\bar\rho_s= m_s/({4\pi\over 3}r_s^3)$, and  one fixed but arbitrary internal datum $(p_*,r_*)$.

Using techniques similar to the above, it is likely that more information could be extracted from this sort of argument.

\clearpage
\section{Conclusions}

In this article we have contributed to the rich tapestry of results known regarding pressure profiles in general relativistic perfect fluid spheres. 
Specifically we have developed a number of (relatively straightforward but sometimes somewhat tedious) explicit bounds on the pressure profiles of general relativistic static perfect fluid spheres. We have carefully varied the strength of the input hypotheses to extract as much general information as possible, and have carefully compared the output results. Even the mildest of the input assumptions on the density profile yields useful output regarding the pressure profile.

 \bigskip
 \hrule\hrule\hrule

\appendix
\section{{Appendix: Pressure derivatives at the origin}}

This appendix is designed for use in Corollary 2b; specifically to lift the degeneracy in this equation:
\begin{equation}
\left. {d^2\hat p(\rho_c,K_c; r)\over dr^2}\right|_0  = \left. {d^2 p(r)\over dr^2}\right|_0 < \left. {d^2\hat p(\rho_s,K_s; r)\over dr^2}\right|_0 .
\end{equation}

Recall that to keep $p(r)$ and $\rho(r)$ finite and differentiable at the centre, we need these quantities to depend only on even powers of $r$:
\begin{equation}
p(r) = p_c +{(p'')_c\over 2!} r^2 + {(p'''')_c\over 4!} r^4 + \O(r^6);
\end{equation}
\begin{equation}
\rho(r) = \rho_c +{(\rho'')_c\over 2!} r^2 + {(\rho'''')_c\over 4!} r^4 + \O(r^6).
\end{equation}
See for instance references~\cite{Visser-Yunes} and~\cite{Arrechea:2026}.

Since for a semi-realistic we want the density to be positive and exhibit a maximum at the centre we should enforce $(\rho'')_c <0$.
This can be done by choosing
\begin{equation}
(\rho'')_c = - {\rho_c\over a^2}; \qquad (a \in (0,\infty]).
\end{equation}
Here $a$ is a physical distance characterizing the scale on which $\rho(r)$ varies.
Then
\begin{equation}
\rho(r) = \rho_c -{\rho_c\over 2} {r^2\over a^2} + \O(r^4);
\end{equation}
implying
\begin{equation}
m(r) = 4\pi\rho_c\left(  {r^3 \over3}- {r^5\over 10 a^2} + \O(r^7)\right).
\end{equation}
That is, based on finiteness, differentiability, and (now in addition) positivity and maximality of $\rho(r)$ at the origin, one has :
\begin{equation}
p(r) = p_c +{(p'')_c\over 2} r^2 + {(p'''')_c\over 4!} r^4 + \O(r^6);
\end{equation}
\begin{equation}
\rho(r) = \rho_c -{\rho_c\over 2} {r^2\over a^2} + \O(r^4);
\end{equation}
\begin{equation}
m(r) = 4\pi \rho_c\left(  {r^3 \over3}- {r^5\over 10 a^2} + \O(r^7)\right).
\end{equation}

Now insert these expansions into the TOV equation expressed in the form
\begin{equation}
{dp(r)\over dr} + {[\rho(r)+p(r)]\;[m(r)+4\pi p(r) r^3] \over r^2[1-2m(r)/r]} = 0.
\end{equation}
Taylor series expand this around $r=0$:
\begin{eqnarray}
&&
\left( (p'')_c + {4\pi\over3} [\rho_c+p_c][\rho_c+3 p_c] \right) r  \hfill
\nonumber
\\
&&+ {1\over 6} \left( (p'''')_c   +(p'')_c 8\pi(2\rho_c + 3 p_c) + {64 \pi^2\over3} \rho_c [\rho_c+p_c][\rho_c + 3 p_c]-{8\pi\over5} {\rho_c(4\rho_c+9p_c)\over 5 a^2}\right) r^3
\nonumber
\\
&&+ \O(r^5) = 0.
\end{eqnarray}
From the term proportional to $r$ we regain one of our previous results
\begin{equation}
(p'')_c = -  {4\pi\over3} [\rho_c+p_c][\rho_c+3 p_c] .
\end{equation}
From the term proportional to $r^3$ we now see
\begin{equation}
(p'''')_c   = - (p'')_c 3\pi(2\rho_c + 3 p_c) - 64 \pi^2 \rho_c [\rho_c+p_c][\rho_c + 3 p_c]-{8\pi\over5} {\rho_c(4\rho_c+9p_c)\over a^2}.
\end{equation}
Using the known expression for $(p'')_c$ we deduce
\begin{equation}
(p'''')_c   =  32 \pi^2 p_c [\rho_c+p_c][\rho_c + 3 p_c]+{8\pi\over5} {\rho_c(4\rho_c+9p_c)\over  a^2}
\end{equation}
For a constant density star $a\to\infty$ and we have
\begin{equation}
\left.(p'''')_c\right|_{{\scriptstyle constant\,density\,star}}   =  32 \pi^2 p_c [\rho_c+p_c][\rho_c + 3 p_c]
\end{equation}
For a generic star with maximal density at the centre $a\in(0,\infty)$, and so we have
\begin{equation}
\left.(p'''')_c\right|_{{\scriptstyle generic\,star}}   > 
\left.(p'''')_c\right|_{{\scriptstyle constant\,density\,star}} 
\end{equation}


Application to Corollary 2-b: 
Instead of merely having
\begin{equation}
\left. {d^2\hat p(\rho_c,K_c; r)\over dr^2}\right|_0  = \left. {d^2 p(r)\over dr^2}\right|_0 < \left. {d^2\hat p(\rho_s,K_s; r)\over dr^2}\right|_0,
\end{equation}
we now see that we have both
\begin{equation}
\left. {d^4\hat p(\rho_c,K_c; r)\over dr^4}\right|_0  < \left. {d^4p(r)\over dr^4}\right|_0
\qquad 
\hbox{and}
\qquad
\left. {d^2 p(r)\over dr^2}\right|_0 < \left. {d^2\hat p(\rho_s,K_s; r)\over dr^2}\right|_0 .
\end{equation}
which completes the justification of
\begin{equation}
\hat p(\rho_c,K_c; r) < p(r) < \hat p(\rho_s,K_s; r).
\end{equation}
Explicitly:
{\small
\begin{eqnarray}
&&
 \rho_c \left( [\rho_c+3 p_c] \sqrt{1-{8\pi\over3} \rho_c r^2} - [\rho_c+ p_c] 
\over 3 [\rho_c+ p_c] -  [\rho_c+3 p_c] \sqrt{1-{8\pi\over3} \rho_c r^2}\right)
\leq p(r) \leq
\rho_s \left( [\rho_s+3 p_c] \sqrt{1-{8\pi\over3} \rho_s r^2} - [\rho_s + p_c]
\over 3 [\rho_s + p_c] -  [\rho_s+3 p_c] \sqrt{1-{8\pi\over3} \rho_s r^2}\right); \nonumber\\
&&
\end{eqnarray}}
This now fully completes all technical steps in deriving Corollary 2b.


\bigskip
\bigskip
\hrule\hrule\hrule

\clearpage
\hrule\hrule\hrule
\addtocontents{toc}{\bigskip\hrule\hrule\hrule}

\vspace{-25pt}
\setcounter{secnumdepth}{0}
\section[\hspace{14pt}  References]{}
%

 \end{document}